# The EU AI Act and the Rights-Based Approach to Technological Governance

Georgios Pavlidis

Associate Professor of International and EU Law, Director of the Jean Monnet Center of Excellence AI-2-TRACE-CRIME, Neapolis University Pafos (NUP), School of Law; correspondence address: Neapolis University Pafos, Office 253, Danaes Avenue 2, Pafos 8042, Cyprus; e-mail: g.pavlidis@nup.ac.cy
https://orcid.org/0000-0001-6311-3086

**Abstract:** The European Union AI Act constitutes an important development in shaping the Union's digital regulatory architecture. The Act places fundamental rights at the heart of a risk-based governance framework. The article examines how the AI Act institutionalizes a human-centric approach to AI and how the AI Act's provisions explicitly and implicitly embed the protection of rights enshrined in the EU Charter of Fundamental Rights. It argues that fundamental rights function not merely as aspirational goals, but as legal thresholds and procedural triggers across the life cycle of an AI system. The analysis suggests that the AI Act has the potential to serve as a model for rights-preserving AI systems, while acknowledging that challenges will emerge at the level of implementation.

## 1. Introduction

The Artificial Intelligence Act (AI Act)[1] of the European Union (EU) is the first comprehensive attempt by a major jurisdiction to regulate artificial intelligence (AI) through a horizontal legal framework, which follows a risk-based approach (RBA).[2] The AI Act was published in the Official Journal of the European Union on July 12, 2024, and entered into force on August 1, 2024. While its general application is scheduled for August 2, 2026, several provisions are subject to phased application, with certain obligations becoming applicable as early as 2025 and others deferred until 2027. The objective of the AI Act is to ensure the development and use of AI systems in alignment with the core values of

This study was funded by the European Union. Views and opinions expressed are, however, those of the author only and do not necessarily reflect those of the European Union or the European Education and Culture Executive Agency (EACEA). Neither the European Union nor EACEA can be held responsible for them.

1 Regulation (EU) 2024/1689 of the European Parliament and of the Council of 13 June 2024 laying down harmonised rules on artificial intelligence (Artificial Intelligence Act) (OJ L2024/1689, 12 July 2024).

2 This approach has been used in several regulatory contexts, such as anti-money laundering; Gauri Sinha, "Risk-Based Approach: Is It the Answer to Effective Anti-Money Laundering Compliance?," in *Assets, Crimes and the State: Innovation in 21st Century Legal Responses*, eds. Katie Benson, Colin King, and Clive Walker (London: Routledge, 2020), 52–65; Georgios Pavlidis, "Asset Recovery in the European Union: Implementing a 'No Safe Haven' Strategy for Illicit Proceeds," *Journal of Money Laundering Control* 25, no. 1 (2022): 109, https://doi.org/10.1108/JMLC-11-2020-0131.

the EU, including the protection of fundamental rights, as guaranteed by the Charter of Fundamental Rights of the European Union (the Charter).[3] Earlier EU initiatives, such as the General Data Protection Regulation (GDPR)[4] or the Digital Services Act (DSA),[5] address specific aspects of data protection or content moderation. However, unlike these initiatives, the AI Act integrates fundamental rights considerations across the entire life cycle of AI systems. The AI Act treats fundamental rights not as ancillary issues, but as part of legal obligations. This article examines how fundamental rights are integrated into the structure and logic of the AI Act. Special attention is given to the dynamic mechanisms, such as conformity assessments and institutional oversight, that are conditioned by the protection of fundamental rights. It is argued that the AI Act employs a model of rights-driven technological governance, using formal prohibitions, risk classifications, and continuous oversight grounded in the Charter.

## 2. The Rights-Based Approach: The Normative Compass of the AI Act

Article 1 of the AI Act sets the tone for the entire Regulation by aligning the development and deployment of AI systems[6] with the protection of fundamental rights, in compliance with the Charter of Fundamental Rights of the European Union. This reflects a strong commitment by the EU to ensure that the advancement of AI does not come at the expense of human dignity, privacy, non-discrimination, and other core rights.[7] Since the AI Act aims to promote "human-centric"[8] and "trustworthy"[9] AI, it acknowledges that technological innovation must serve people, not override their freedoms. Article 1 thus functions as both a declaratory provision and a normative compass, ensuring that innovation does not sideline fundamental rights, but instead reinforces them in the digital era. In practical terms, this means that all subsequent rules of the AI Act must be interpreted through the lens of fundamental rights.

The territorial, personal, and material scope of application of the AI Act reinforces the protection of fundamental rights by adopting an expansive approach. Indeed, Article 2(1)(c) ensures that the AI Act applies extraterritorially: even if AI providers[10] or

---

3 Charter of Fundamental Rights of the European Union (OJ C326/391, 26 October 2012).

4 Regulation (EU) 2016/679 of the European Parliament and of the Council of 27 April 2016 on the protection of natural persons with regard to the processing of personal data and on the free movement of such data, and repealing Directive 95/46/EC (General Data Protection Regulation)(OJ L119/1, 4 May 2016).

5 Regulation (EU) 2022/2065 of the European Parliament and of the Council of 19 October 2022 on a Single Market for Digital Services (Digital Services Act)(OJ L277/1, 27 October 2022).

6 According to Article 3(1) AI Act, "AI system" means a machine-based system that is designed to operate with varying levels of autonomy and that may exhibit adaptiveness after deployment, and that, for explicit or implicit objectives, infers, from the input it receives, how to generate outputs such as predictions, content, recommendations, or decisions that can influence physical or virtual environments.

7 Recitals 1 and 2 AI Act.

8 Joanna Bryson and Andreas Theodorou, "How Society Can Maintain Human-Centric Artificial Intelligence," in *Human-Centered Digitalization and Services*, eds. Marja Toivonen and Eveliina Saari (Singapore: Springer, 2019), 305–23.

9 Luciano Floridi, *Ethics, Governance, and Policies in Artificial Intelligence* (Cham: Springer, 2021), 41–5.

10 Under Article 3(3) AI Act, "provider" means a natural or legal person, public authority, agency or other body that develops an AI system or a general-purpose AI model or that has an AI system or a general-purpose AI deployers[11] are established outside the EU, they fall within scope when their AI systems produce outputs that are used within the Union. This will help mitigate external threats, such as discriminatory or manipulative AI systems deployed transnationally.[12] Article 2(1)(g) makes it clear that the AI Act applies when people affected by an AI system are located within the EU. This puts individuals and their rights at the heart of the Regulation's scope, in line with the Charter's focus on personal dignity and protection. For its part, Article 2(4) adds an important nuance: it allows certain public authorities and international organizations to be exempt from the Act when they are working within the context of international cooperation in law enforcement or judicial matters. However, this exemption only applies if there are adequate safeguards in place to protect fundamental rights and freedoms, which is a recurrent concern in international police or judicial cooperation.[13] Such safeguards are important in cross-border contexts,[14] which can be prone to weaker oversight.

The AI Act defines key concepts that shape the interpretation and application of the rules on AI governance, including important references to fundamental rights. Two definitions stand out for their explicit rights-based dimension. First, the definition of a "serious incident" under Article 3, point (49) of the AI Act includes not only physical harm or disruption to infrastructure, but also incidents that result in "the infringement of obligations under Union law intended to protect fundamental rights."[15] Thus, breaches of rights, such as violations of privacy, non-discrimination, or due process, are treated with the same gravity as harms to human life and health, property, and infrastructure. This reflects an understanding that the harms caused by AI systems are not limited to physical consequences, but include insidious or systemic interferences with fundamental rights.[16] Second, the definition of "systemic risk" in Article 3, point (65), extends the notion of risk to include negative effects on fundamental rights and society as a whole, particularly from general-purpose AI models with high-impact capabilities. Therefore, these rights-sensitive concepts are integrated at the definitional level, affecting the interpretation of all subsequent provisions.

Another protective function of the AI Act is to categorically prohibit certain AI practices because they are deemed incompatible with the values and fundamental rights model developed and places it on the market or puts the AI system into service under its own name or trademark, whether for payment or free of charge.

11 Under Article 3(4) AI Act, "deployer" means a natural or legal person, public authority, agency, or other body using an AI system under its authority except where the AI system is used in the course of a personal non-professional activity.

12 Huw Roberts et al., "Global AI Governance: Barriers and Pathways Forward," *International Affairs* 100, no. 3 (2024): 1275, https://doi.org/10.1093/ia/iiae073.

13 Giulio Calcara, "Balancing International Police Cooperation: INTERPOL and the Undesirable Trade-off Between Rights of Individuals and Global Security," *Liverpool Law Review* 42, no. 2 (2021): 111, https://doi.org/10.1007/s10991-020-09266-9.

14 Recital 3 AI Act.

15 Recital 155 AI Act.

16 Francesca Palmiotto, "The AI Act Roller Coaster: The Evolution of Fundamental Rights Protection in the Legislative Process and the Future of the Regulation," *European Journal of Risk Regulation* 16, no. 2 (2025): 770–93, https://doi.org/10.1017/err.2024.97.

under the EU Charter.[17] Among these practices, Article 5(1)(h) addresses a very controversial application of AI: "real-time" remote biometric identification systems[18] in publicly accessible spaces for law enforcement purposes. The default prohibition of such practices reflects serious concerns over their potential to undermine rights to privacy and data protection (Articles 7 and 8 of the Charter), freedom of assembly (Article 12), and non-discrimination (Article 21), as well as broader risks to democratic participation.[19] This preventive logic echoes the Court of Justice's reasoning in *Digital Rights Ireland*, where it held that large-scale, indiscriminate technological interferences with privacy and data protection are incompatible with the Charter, unless strictly necessary and proportionate, thereby reinforcing the rights-based limits on biometric surveillance.[20] The legislative text also introduces some narrowly tailored exceptions, which apply to exceptional circumstances, such as a targeted search for specific victims of abduction, trafficking in human beings, or sexual exploitation of human beings, the search for missing persons, as well as the prevention of imminent threats to life, or a "present, or genuine and foreseeable, threat of a terrorist attack," or the pursuit of serious criminal offences.[21] Even then, Article 5(2) requires a proportionality assessment that is context-sensitive. Furthermore, the use of such systems by law enforcement authorities is conditioned on a fundamental rights impact assessment (FRIA)[22] under Article 27 and prior registration in the EU database (Article 49).[23] These conditions promote accountability in the use of AI systems for "real-time" remote biometric identification. Real time biometric surveillance, even when exceptionally permitted, must be treated as a last resort, bounded by legal safeguards and the principle of necessity.[24] The objective is to protect the public sphere and curtail techno-authoritarian tools and indiscriminate biometric surveillance in a pre-emptive manner.[25]

## 3. The Rights-Based Approach in the Case of High-Risk AI Systems

In addition to the rules on prohibited AI systems, the AI Act introduces the classification of "high-risk" AI systems, triggering stringent obligations. Article 6, read in

17 Rostam J. Neuwirth, "Prohibited Artificial Intelligence Practices in the Proposed EU Artificial Intelligence Act (AIA)," *Computer Law & Security Review* 48 (2023): 105798, https://doi.org/10.1016/j.clsr.2023.105798.

18 Article 3 (41) and (42) AI Act.

19 Recitals 32 ff AI Act.

20 CJEU Judgment of 8 April 2014, Digital Rights Ireland Ltd v. Minister for Communications, Marine and Natural Resources and Others, Joined Cases C-293/12 and C-594/12, ECLI:EU:C:2014:238.

21 Annex II AI Act.

22 Recital 96 AI Act.

23 Recital 131 AI Act.

24 Arvind Jaiswal and Sandhya Tarar, "Real-Time Biometric System for Security and Surveillance Using Face Recognition," in *Advances in Computing and Data Sciences: 4th International Conference, ICACDS 2020, Valletta, Malta, April 24–25, 2020, Revised Selected Papers*, eds. Mayank Singh et al. (Singapore: Springer, 2020), 293–304, https://doi.org/10.1007/978-981-15-6634-9.

25 Hendrik Schopmans and İrem Tuncer Ebetürk, "Techno-Authoritarian Imaginaries and the Politics of Resistance Against Facial Recognition Technology in the US and European Union," *Democratization* 31, no. 5 (2024): 943–62, https://doi.org/10.1080/13510347.2023.2258803.

conjunction with Annex III, defines the criteria for such designations. Categories of "high-risk" AI systems include biometrics, critical infrastructure, education and vocational training, employment, workers' management and access to self-employment, access to and enjoyment of essential private services and essential public services and benefits, law enforcement, migration, asylum and border control management, administration of justice, and democratic processes.[26] An AI system listed in Annex III is presumed to be high-risk, unless it demonstrably does not pose a significant risk to health, safety, or fundamental rights. Therefore, the AI Act introduces a threshold for certain categories of systems that may threaten fundamental rights, although it could be argued that the standard of "significant risk to fundamental rights" leaves room for uncertainty and needs detailed interpretive guidance. Article 6(3) also introduces certain narrowly defined conditions under which an AI system may escape the high-risk classification, where it does not pose a significant risk of harm to the health, safety, or fundamental rights of natural persons, including by not materially influencing the outcome of decision-making. In this context, the AI system must be intended to perform a narrow procedural task or improve the result of a previously completed human activity; detect decision-making patterns or deviations from prior decision-making patterns and is not meant to replace or influence the previously completed human assessment without proper human review; or perform a preparatory task for an assessment relevant for the purposes of the use cases listed in Annex III. The exemptions under Article 6 are carefully delimited, and they do not apply if the AI system performs profiling of natural persons.[27] Such profiling inherently raises serious fundamental rights concerns, including the rights to privacy, dignity, and equality.[28]

Another important compliance obligation concerns the establishment of risk management systems for high-risk AI systems. [29] Such systems, grounded in Article 9, must encompass risks to fundamental rights, alongside risks to health and safety.[30] This is another example of the AI Act's rights-based approach. Risks to fundamental rights, such as privacy, non-discrimination, due process, and freedom of expression, must be identified, analyzed, evaluated, and mitigated throughout the entire life cycle of the AI system. Providers are mandated to assess not only known risks, but also reasonably foreseeable risks to fundamental rights. This compels proactive engagement with how the AI systems might impact individuals and groups in the real world, including in ways not initially intended. Reasonably foreseeable misuses[31] must be avoided, because fundamental rights can be violated not only by flawed design, but also by predictable deployment patterns, such as biased training data. Moreover, there is a requirement to

26 Ali Sunyaev et al., "High-Risk Artificial Intelligence," *Business & Information Systems Engineering* 67 (2025): 981, https://doi.org/10.1007/s12599-025-00942-6.

27 The term "profiling" is defined in Article 4, point (4) of GDPR.

28 Laurie N. Hobart, "AI, Bias, and National Security Profiling," *Berkeley Technology Law Journal* 40, no. 1 (2025): 165–231, https://doi.org/10.15779/Z38VX06474.

29 Jonas Schuett, "Risk Management in the Artificial Intelligence Act," *European Journal of Risk Regulation* 15, no. 2 (2024): 367–85, https://doi.org/10.1017/err.2023.1.

30 Recital 67 AI Act.

31 Article 3(13) AI Act.

integrate insights from post-market monitoring systems,[32] which creates a feedback loop and enhances accountability.

In addition to risk management systems, the AI Act introduces another safeguard: the protection of fundamental rights through data governance.[33] Data is the foundation on which AI systems are trained and perform.[34] Therefore, flawed or biased data can translate into discriminatory outcomes, violations of privacy, and other systemic injustices.[35] High-risk AI systems must be developed using training, validation, and testing data sets that meet stringent criteria, precisely to prevent such harms. Among these criteria, Article 10(2)(f) explicitly requires examination of data sets for biases that are likely to affect health and safety, or have a negative impact on fundamental rights.

## 4. Procedural and Substantive Safeguards for the Protection of Fundamental Rights

### 4.1. Transparency Obligations and Fundamental Rights

The AI Act recognizes the significant role of transparency in protecting fundamental rights.[36] Article 13 specifically requires that the design of high-risk AI systems must enable deployers to understand and appropriately use the AI outputs. This is important, since deployers may not have the necessary technical expertise to identify and mitigate risks to privacy, non-discrimination, due process, or access to services. Transparency becomes a design feature of high-risk AI systems, rather than a fix that is applied *post hoc*.[37] This is a preventive approach that integrates the awareness of fundamental rights into the architecture of high-risk AI systems.[38]

The AI Act explicitly links transparency to the protection of fundamental rights, and it requires providers to disclose any foreseeable risks to health, safety, or fundamental rights resulting from both intended use and reasonably foreseeable misuse.[39] This is another anticipatory obligation, as used in other sections of the Act, which empowers deployers to assess the real-world, human impact of the AI system and adjust their practices accordingly.[40] There are additional requirements, such as providing information relevant to explainability, performance across demographic groups, and data specifications.[41] These requirements are tied to the right to non-discrimination and equal treatment, especially where the AI outputs may vary across populations. More broadly, transparency obligations under Article 13 serve as a safeguard for the right to good administration and the right to an effective remedy and to a fair trial (Articles 41 and 47 of the Charter).[42] Indeed, access to clear, comprehensible system documentation is a prerequisite for individuals to take legal action and contest adverse decisions based on AI outputs. In this respect, the emphasis on intelligibility and effective contestation aligns with the Court of Justice's approach in *Schrems II*, which stressed that formal legal safeguards are insufficient where individuals lack practical means to understand, challenge, or obtain redress against data-processing operations.[43] Meaningful transparency is important for affected persons, as it ensures that the logic behind decisions is not a "black box"[44] for the individuals whose rights are at stake.

---

32 Jakob Mökander et al., "Conformity Assessments and Post-Market Monitoring: A Guide to the Role of Auditing in the Proposed European AI Regulation," *Minds and Machines* 32, no. 2 (2022): 241–68, https://doi.org/10.1007/s11023-021-09577-4.

33 Marijn Janssen et al., "Data Governance: Organizing Data for Trustworthy Artificial Intelligence," *Government Information Quarterly* 37, no. 3 (2020): 101493, https://doi.org/10.1016/j.giq.2020.101493.

34 Bogdan Fischer and Agnieszka Piskorz-Ryń, "Artificial Intelligence in the Context of Data Governance," *International Review of Law, Computers & Technology* 35, no. 3 (2021): 419–28, https://doi.org/10.1080/13600869.2021.1950925.

35 Recital 66 AI Act.

36 Nagadivya Balasubramaniam et al., "Transparency and Explainability of AI Systems: From Ethical Guidelines to Requirements," *Information and Software Technology* 159 (2023): 107197, https://doi.org/10.1016/j.infsof.2023.107197.

37 Heike Felzmann et al., "Towards Transparency by Design for Artificial Intelligence," *Science and Engineering Ethics* 26, no. 6 (2020): 3333, https://doi.org/10.1007/s11948-020-00276-4.

38 Ognyan Seizov and Alexander J. Wulf, "Artificial Intelligence and Transparency: A Blueprint for Improving the Regulation of AI Applications in the EU," *European Business Law Review* 31, no. 4 (2020): 611–40, https://doi.org/10.54648/eulr2020024; Recital 72 AI Act.

39 Article 13(3) point (b)(iii) AI Act.

### 4.2. Human Oversight as a Rights-Preserving Principle

Human oversight is another important safeguard in the governance of high-risk AI systems.[45] The AI Act introduces such oversight in its Article 14, as a mechanism to protect fundamental rights, alongside health and safety. The objective is that AI remains accountable to human judgment and does not operate in a normative vacuum.[46] This is particularly important where algorithmic outputs significantly affect people's lives, whether in policing, hiring, migration control, education, or access to welfare. Therefore, AI systems must be designed and developed in a way that allows oversight by natural persons. The AI Act affirms the principle that human agency must not be displaced by AI when decision-making can adversely affect individual rights and dignity. Oversight under Article 14(2) is essential for preventing or minimizing risks to fundamental rights, not just under intended use but also under reasonably foreseeable misuse. Oversight must be effective, not just symbolic. A similar concern was articulated by the European Court of Human Rights in *Big Brother Watch*, where the Court emphasized that human

40 Alessandro Mantelero, "AI and Big Data: A Blueprint for a Human Rights, Social and Ethical Impact Assessment," *Computer Law & Security Review* 34, no. 4 (2018): 754–72, https://doi.org/10.1016/j.clsr.2018.05.017.

41 Article 13(3) points (b)(iv) to (b)(vii) AI Act.

42 Kathleen Gutman, "The Essence of the Fundamental Right to an Effective Remedy and to a Fair Trial in the Case-Law of the Court of Justice of the European Union: The Best Is Yet to Come?," *German Law Journal* 20, no. 6 (2019): 884–903, https://doi.org/10.1017/glj.2019.67; Izabela M. Wróbel, "Artificial Intelligence Systems and the Right to Good Administration," *Review of European and Comparative Law* 49, no. 2 (2022): 203–23, https://doi.org/10.31743/recl.13616.

43 CJEU Judgment of 16 July 2020, Data Protection Commissioner v. Facebook Ireland Ltd, Maximillian Schrems, Case C-311/18, ECLI:EU:C:2020:559.

44 Georgios Pavlidis, "Unlocking the Black Box: Analysing the EU Artificial Intelligence Act's Framework for Explainability in AI," *Law, Innovation and Technology* 16, no. 1 (2024): 293–308, https://doi.org/10.1080/17579961.2024.2313795.

45 Riikka Koulu, "Proceduralizing Control and Discretion: Human Oversight in Artificial Intelligence Policy," *Maastricht Journal of European and Comparative Law* 27, no. 6 (2020): 720–35, https://doi.org/10.1177/1023263X20978649.

46 Recital 73 AI Act.

oversight mechanisms must be capable of providing real and continuous control over automated or large-scale surveillance systems, rather than operating as purely formal or *ex post* safeguards.[47] This concept anticipates deployment in real-world scenarios where discrimination and other violations of fundamental rights may arise unintentionally or through negligence. Of course, challenges remain regarding scalability, since oversight models may not scale effectively for high-volume, automated decisions. The AI Act gives some flexibility in how oversight is implemented. First, human oversight measures can be identified and built, when technically feasible, into the high-risk AI system (e.g., real-time alerts, override functions) by the provider before it is placed on the market or put into service. Second, such measures may be identified by the provider before placing the high-risk AI system on the market or putting it into service, which must be appropriate for implementation by the deployer. In both cases, the measures must be proportionate and context-sensitive. AI must not function as an unreviewable authority, and AI outputs must be contestable and corrigible by humans when fundamental rights are at stake. Thus, the AI Act gives practical effect to emerging principles, such as human-in-the-loop and human-on-the-loop,[48] taking into consideration the protection of fundamental rights, as discussed above.

### 4.3. Fundamental Rights Impact Assessment for High-Risk AI Systems

The AI Act introduces a significant procedural innovation: the mandatory Fundamental Rights Impact Assessment (FRIA) for certain deployers of high-risk AI systems.[49] An FRIA, as required under Article 27, deals with the use of AI in specific, sensitive contexts, such as public administration, law enforcement, and essential service provision, which can affect individuals' rights to privacy, non-discrimination, due process, education, and social protection. The AI Act requires public sector bodies and private entities performing public functions to assess such risks before the deployment of AI tools in these contexts. The nature of this accountability mechanism under Article 27 is clearly preventive.[50] The approach to the development of FRIAs is granular. Deployers must consider not only the system's technical characteristics, but also the context of its use in the real world. They are also required to consider the categories of individuals affected and the potential for harmful or discriminatory outcomes. This aligns with the Charter's focus on vulnerability, equality, and human dignity because deployers must assess specific risks to different groups, not just the general population. The AI Act further requires the FRIA to describe oversight mechanisms and redress arrangements, which is associated with the right to an effective remedy.[51] Deployers are allowed to rely on prior assessments, including assessments conducted by the provider, but such assessments must be kept up to date. The FRIA must also be notified to market surveillance authorities using a recordable and reviewable interface between AI deployment and public oversight. Finally, the FRIA complements, rather than duplicates, the Data Protection Impact Assessment (DPIA) required under the GDPR.[52] Thus, privacy and broader rights considerations are integrated into a unified compliance framework.[53] Nevertheless, it must be taken into account that FRIAs have limited applicability, in that they are binding only for specific categories of deployers, while practical guidance on their implementation remains under development. As a result, the FRIA functions not as a universal, *ex ante* obligation across all high-risk AI deployments, but as a targeted safeguard whose scope and operational content depend on both the classification of the system and the institutional role of the deploying actor.

[47] ECtHR Judgment of 25 May 2021, Big Brother Watch and Others v. United Kingdom, application nos. 58170/13, 62322/14 and 24960/15.

[48] Therese Enarsson, Lena Enqvist, and Markus Naarttijärvi, "Approaching the Human in the Loop–Legal Perspectives on Hybrid Human/Algorithmic Decision-Making in Three Contexts," *Information & Communications Technology Law* 31 (2022): 123–53, https://doi.org/10.1080/13600834.2021.1958860.

[49] Recital 96 AI Act.

[50] Alessandro Mantelero, "The Fundamental Rights Impact Assessment (FRIA) in the AI Act: Roots, Legal Obligations and Key Elements for a Model Template," *Computer Law & Security Review* 54 (2024): 106020, https://doi.org/10.1016/j.clsr.2024.106020.

### 4.4. Right to Explanation of Individual Decision-Making

The AI Act explicitly recognizes a right to explanation for individuals affected by decisions based on the output of high-risk AI systems.[54] This right, protected in Article 86, follows the logic of procedural fairness, transparency, and accountability. It operates as a prerequisite for the effective protection of fundamental rights, particularly in high-stakes contexts where algorithmic outputs influence access to services or opportunities in areas such as education, employment, credit, migration, and justice. At the same time, the right to explanation under the AI Act is conditional in scope and does not amount to a general or absolute entitlement applicable to all AI systems, but is instead linked to specific regulatory contexts, system classifications, and the nature of the decision-making process at issue. The right to explanation covers decisions taken by a deployer of a high-risk AI system (Annex III) that produce legal effects or similarly significant consequences. The right is triggered when the individual perceives the decision as having an adverse impact on their health, safety, or fundamental rights. This links the right to explanation to the implementation of the EU Charter, notably Articles 41 (right to good administration), 47 (right to an effective remedy), and 8 (protection of personal data). What makes Article 86 particularly impactful is that it requires not just a technical disclosure, but a "clear and meaningful explanation" of the following: (1) the role played by the AI system in the decision-making process, and (2) the main elements of the decision itself. The objective is to prevent "black-box" decisions that are unintelligible to the average person.[55] Such opaque decisions may impede access to redress or judicial review, undermining transparency and contestability, as discussed above. However, there are important limitations to the right to explanation, which does not apply where lawful exceptions are provided under Union or national law, and it only applies in the absence of similar rights already guaranteed elsewhere in EU law.[56] These exceptions are narrowly drawn and subject to Union law compliance. For this reason, they do not limit the importance of Article 86, which reflects the broader principles of human-centric, accountable, and contestable AI. In any case, it must be noted that, despite the right to explanation, the practical ability to challenge AI decisions remains rather constrained by information asymmetries and legal complexity.

---

51 Angela Ward, "Remedies Under the EU Charter of Fundamental Rights," in *Research Handbook on EU Law and Human Rights*, eds. Sionaidh Douglas-Scott and Nicholas Hatzis (Cheltenham: Edward Elgar Publishing, 2017), 162–85.

52 Katerina Demetzou, "Data Protection Impact Assessment: A Tool for Accountability and the Unclarified Concept of 'High Risk' in the General Data Protection Regulation," *Computer Law & Security Review* 35, no. 6 (2019): 105342, https://doi.org/10.1016/j.clsr.2019.105342.

53 Article 27(4) AI Act.

54 Fleur Jongepier and Esther Keymolen, "Explanation and Agency: Exploring the Normative-Epistemic Landscape of the 'Right to Explanation,'" *Ethics and Information Technology* 24, no. 49 (2022), https://doi.org/10.1007/s10676-022-09654-x.

## 5. Oversight and Enforcement through the Lens of Fundamental Rights

### 5.1. Tasks of the European Artificial Intelligence Board

The AI Act delineates the tasks of the European Artificial Intelligence Board (the Board), which functions as a coordinating and advisory body for the application of the AI Act across the Union. The protection of fundamental rights is a recurrent theme in the mandate of the Board.[57] This is most evident in several core responsibilities. First, the Board is tasked with facilitating coordination, harmonization of administrative practices, and the dissemination of best practices, including with respect to AI regulatory sandboxes, which themselves must mitigate fundamental rights risks, as discussed above. Moreover, the Board may issue recommendations and opinions on matters critical to fundamental rights governance, including revisions to Annex III (high-risk use cases) and Article 5 (prohibited practices), discussed above. The Board is also empowered to support AI literacy, public awareness, and understanding of safeguards and rights. The Board is also tasked with cooperating with EU agencies and networks in domains such as data and fundamental rights protection. Thus, the Board can function as a governance node that connects national authorities, the AI Office, the Commission, and broader civil society.[58] It is worth mentioning that the AI Act also establishes the Advisory Forum, a permanent, multistakeholder consultative body, to support the European AI Board and the Commission by offering technical expertise and strategic guidance.[59] The EU Agency for Fundamental Rights (FRA)[60] has been designated as a permanent member of the Advisory Forum, which guarantees a continuous and expert voice on fundamental rights in the

55 Margot E. Kaminski, "The Right to Explanation, Explained," in *Research Handbook on Information Law and Governance*, eds. Sharon Sandeen, Christoph Rademacher, and Ansgar Ohly (Cheltenham: Edward Elgar Publishing, 2021), 278–99.

56 For example, see Articles 15 or 22 GDPR.

57 Article 66 AI Act.

58 Claudio Novelli et al., "A Robust Governance for the AI Act: AI Office, AI Board, Scientific Panel, and National Authorities," *European Journal of Risk Regulation* 16, no. 2 (2025): 566–90, https://doi.org/10.1017/err.2024.57.

59 Article 67 AI Act.

60 Council Regulation (EC) No 168/2007 of 15 February 2007 establishing a European Union Agency for Fundamental Rights (OJ L53/1, 22 February 2007).

discussions surrounding AI governance. The inclusion of ENISA[61] (cybersecurity) and the main European standardization bodies (CEN, CENELEC, and ETSI) further strengthens the role and expertise of the Forum at the intersection of rights, safety, and technical standards.

### 5.2. Designation of National Competent Authorities

At the level of enforcement, the AI Act requires each Member State to designate national competent authorities[62] and a single point of contact. Article 70(3) of the AI Act explicitly integrates fundamental rights protection into this institutional design; it requires that the designated authorities possess expertise not only in AI technologies, but also in personal data protection and fundamental rights. Therefore, the authorities tasked with monitoring and enforcing the AI Act cannot be merely technical regulators but must also be equipped to assess the legal and ethical implications of AI systems. Moreover, national competent authorities must act independently, impartially, and without bias, which aligns with the principles of institutional neutrality and accountability. This ensures the legitimacy of decisions taken by national authorities, especially when they deal with sensitive issues related to infringements of fundamental rights. Contact information for these authorities must be publicly available, enabling direct communication with affected persons, including for the lodging of complaints or access to redress mechanisms. Member States must also provide adequate resources and conduct regular assessments of authority competence and capacity to ensure that enforcement is operationally viable. Therefore, Member States must not merely transpose the AI Act formally, but actively build institutional ecosystems that can mitigate risks to fundamental rights across the AI life cycle. The key implementation challenge in this context relates to the diverging institutional capacities between Member States, which may lead to fragmented enforcement and uneven protection of fundamental rights.

### 5.3. Powers of Authorities Protecting Fundamental Rights

The AI Act establishes procedural mechanisms that empower national authorities tasked with enforcing fundamental rights, such as equality bodies, data protection authorities, or human rights institutions, to participate in the oversight of high-risk AI systems. Article 77 gives these authorities a legal basis to access all relevant documentation under the AI Act (risk assessments, technical specifications, post-market monitoring reports, and FRIAs). This must be provided in accessible language and format. The right of access to such documentation allows fundamental rights bodies to assess whether systems deployed in sensitive areas (Annex III) comply not only with the AI Act, but also with the Charter. Where documentation is insufficient, the authorities protecting fundamental rights can request

61 Regulation (EU) 2019/881 of the European Parliament and of the Council of 17 April 2019 on ENISA (the European Union Agency for Cybersecurity) and on information and communications technology cybersecurity certification and repealing Regulation (EU) No 526/2013 (Cybersecurity Act)(OJ L151/15, 7 June 2019).

62 Emanuele Parisini and Eduard Dervishaj, "Emerging Models of National Competent Authorities Under the EU AI Act," *Annual International Conference on Digital Government Research* 26 (2025): 1, https://doi.org/10.59490/dgo.2025.1007.

technical testing of the AI system. This can uncover hidden biases or systemic effects that may not be apparent from documentation alone. Such testing must be carried out in close cooperation with the requesting authority. This is an example of a co-governance model, where both technical regulators and rights enforcers work together.[63]

### 5.4. Procedure at National Level for Dealing with AI Systems Presenting a Risk

The AI Act is integrated within the broader EU product safety regime.[64] The notion of a "product presenting a risk," under Article 79 of the AI Act, includes not only threats to health and safety, but also to fundamental rights. This constitutes an expansion of the traditional understanding of product risk. Market surveillance authorities can actively evaluate AI systems when they have reason to believe such systems present a risk, with particular attention to vulnerable groups. Indeed, the impacts of AI systems may be unevenly distributed, and certain groups, such as children, persons with disabilities, or migrants, may be disproportionately affected. When risks to fundamental rights are identified, the market surveillance authority must inform and cooperate fully with the relevant rights-protecting authorities. An enforcement leverage is ensured because the authorities have the power to require corrective actions, market withdrawal, or recall within short timelines.

### 5.5. AI Regulatory Sandboxes and Fundamental Rights

The AI Act establishes a framework for AI regulatory sandboxes, which are structured environments that enable the development and testing of innovative AI systems under regulatory supervision.[65] These sandboxes aim to promote innovation and market access, especially for SMEs and start-ups.[66] However, the protection of fundamental rights constitutes a condition of participation in these regulatory sandboxes, under Article 57 of the AI Act. This protective function is most evident in Article 57(6), which requires competent authorities supervising sandboxes to provide support not only on technical compliance, but also in identifying, assessing, and mitigating risks to fundamental rights. The inclusion of fundamental rights alongside health and safety throughout Article 57 confirms that innovation within the sandbox must respect specific boundaries. Article 57(11) further strengthens this precautionary approach by requiring authorities to suspend testing where significant rights risks cannot be effectively mitigated. Moreover, Article 57 integrates fundamental rights supervision into experimental AI development. Indeed, national data protection authorities and other relevant regulators must be associated with the operation and oversight of sandboxes where personal data or sectoral rights are involved. This ensures cross-disciplinary oversight. Thus, human rights expertise can be built into

63 Celso Cancela-Outeda, "The EU's AI Act: A Framework for Collaborative Governance," *Internet of Things* 27 (2024): 101291, https://doi.org/10.1016/j.iot.2024.101291.

64 Regulation (EU) 2019/1020 of the European Parliament and of the Council of 20 June 2019 on market surveillance and compliance of products (OJ L169/1, 25 June 2019).

65 Article 3(55) AI Act.

66 Thomas Buocz, Sebastian Pfotenhauer, and Iris Eisenberger, "Regulatory Sandboxes in the AI Act: Reconciling Innovation and Safety?," *Law, Innovation and Technology* 15, no. 2 (2023): 357–89, https://doi.org/10.1080/17579961.2023.2245678.

AI systems from the start, not added as an afterthought once the system is already in use. It must also be noted that participation in a regulatory sandbox does not entail blanket regulatory relief. While Article 57(12) of the AI Act allows for the mitigation of administrative fines where participants act in good faith and in compliance with the agreed sandbox plan and with the guidance of the competent authority, liability under EU and national law remains unaffected, thereby preserving victims' access to judicial and administrative redress. There is also emphasis on transparency and public accountability through exit reports, annual public reporting, and a dedicated interface for stakeholder engagement. All these measures comply with the broader, rights-based governance logic of the AI Act. Thus, regulatory sandboxes can become testbeds for responsible AI, ensuring that innovation remains aligned with fundamental rights. Of course, sandbox operations will depend significantly on the discretion of national authorities, which demonstrates the need for uniform, EU-wide standards in this context.

## 6. Conclusions

The AI Act seeks to answer how emerging technologies can be effectively regulated within liberal, democratic societies.[67] As demonstrated in this article, the protection of fundamental rights is integrated into several key provisions of the AI Act. The Charter is treated not merely as a guiding principle, but as a binding standard for AI governance. Several additional insights emerge from our analysis. First, fundamental rights serve as triggers for regulatory intervention (e.g., in determining high-risk systems), procedural obligations (e.g., transparency and human oversight), and enforcement actions (e.g., responses to compliant but harmful AI systems). Second, the AI Act introduces a dynamic model of governance, whereby the classification of AI systems, the obligations of providers and deployers, and the responsibilities of authorities can be calibrated to address new risks to fundamental rights. This constitutes a move away from static compliance models.

Although this article does not undertake a systematic, comparative analysis, the EU AI Act's rights-based regulatory architecture provides a meaningful reference point for transnational debates on AI governance. The EU model embeds fundamental rights as operative legal thresholds throughout the AI life cycle via risk classification, procedural safeguards, and institutional oversight. Thus, it contrasts with approaches that rely primarily on soft-law standards, sectoral guidance, or *ex post* accountability mechanisms. As such, the EU model offers a structured benchmark against which other emerging regulatory frameworks may be assessed, both within the Union and beyond, particularly in terms of how effectively they translate abstract rights commitments into enforceable procedural guarantees.

Of course, the effectiveness of the rights-based approach of the AI Act will depend on implementation. Translating the Act's provisions into meaningful safeguards requires not only technical expertise, but also institutional capacity, inter-agency cooperation, and civic engagement. Indeed, there are concerns about the operational readiness of national

67 Andreas Jungherr, "Artificial Intelligence and Democracy: A Conceptual Framework," *Social Media + Society* 9, no. 3 (2023), https://doi.org/10.1177/20563051231186353.

authorities, the consistency of conformity assessments, and the meaningful participation of civil society and affected individuals. It is important that future amendments to the AI Act establish additional avenues for public input, redress, or meaningful contestation of high-risk system deployment. Moreover, tensions between innovation and regulation, between business competitiveness and rights protection, will continue to exist. Nonetheless, the AI Act sets an important precedent for how fundamental rights can be preserved in the digital age without retreating from technological change.